\documentclass[final,3p,times,number,sort&compress]{elsarticle}

\usepackage{amssymb}
\usepackage{amsthm}
\usepackage{amsmath}
\usepackage{url}
\usepackage{cases}
\usepackage{booktabs}

\journal{Physica A}

\begin{document}

\begin{frontmatter}



\title{
Influence of self-disassembly of bridges on collective flow characteristics of swarm robots in a single-lane and periodic system with a gap
}


\author[label1]{Kotaro Ito}
\address[label1]{Department of Engineering, Graduate School of Sustainability Science, Tottori University, 4-101 Koyama-cho Minami, Tottori 680-8552, Japan.}
\author[label2]{Ryosuke Nishi\corref{cor2}}\ead{nishi@tottori-u.ac.jp}
\cortext[cor2]{Corresponding author. Tel.: +81 857 31 5192; fax: +81 857 31 5210.}
\address[label2]{Department of Mechanical and Physical Engineering, Faculty of Engineering, Tottori University, 4-101 Koyama-cho Minami, Tottori 680-8552, Japan.}

\begin{abstract}
Inspired by the living bridges formed by ants, swarm robots have been developed to self-assemble bridges to span gaps and self-disassemble them. Self-disassembly of bridges (SDB) may increase the transport efficiency of swarm robots by increasing the number of moving robots, and also may decrease the efficiency by causing gaps to reappear. Our aim is to elucidate the influence of SDB on the collective flow characteristics of swarm robots in a single-lane and periodic system with a gap. In the system, robots span and cross the gap by self-assembling a single-layer bridge. We consider two scenarios in which SDB is prevented (prevent-scenario) or allowed (allow-scenario). We represent the horizontal movement of robots with a typical car-following model, and simply model the actions of robots for self-assembling and self-disassembling bridges. Numerical simulations have revealed the following results. Flow-density diagrams in both the scenarios shift to the higher-density region as the gap length increases. When density is low, allow-scenario exhibits the steady state of repeated self-assembly and self-disassembly of bridges. If density is extremely low, flow in this state is greater than flow in prevent-scenario owing to the increase in the number of robots moving horizontally. Otherwise, flow in this state is smaller than flow in prevent-scenario. Besides, flow in this state increases monotonically with respect to the velocity of robots in joining and leaving bridges. Thus, SDB is recommended for only extremely low-density conditions in periodic systems. Moreover, we have found hysteresis under the absence of periodic boundary conditions (in an open system). This study contributes to the development of the collective dynamics of self-driven particles that self-assemble structures, and stirs the dynamics with other self-assembled structures, such as ramps, chains, and towers.
\end{abstract}


\end{frontmatter}


\section{\label{sec:introduction}Introduction}
Swarm robotics treats a large number of robots that locally interact with other robots and the environment, and aims to achieve robust, scalable, and flexible collective behaviors of robots emerging from the interactions~\citep{Sahin2005}.
Swarm robots realize various collective behaviors that are categorized into four types: spatially organizing behaviors (e.g., self-assembly), navigation behaviors (e.g., collective exploration), collective decision-making (e.g., consensus achievement), and other collective behaviors (e.g., collective fault detection)~\citep{Brambilla2013a}.

When robots explore an unfamiliar field, they may be faced with obstacles they can not go through, such as steps and gaps. Robots overcome the obstacles by the following ways.
In the first way, robots pass over the obstacles without modifying the environment by physically jointing with each other~\citep{Hirose1996, Brown2002, Mondada2004, OGrady2010a, Casarez2016} or mutually transporting each other~\citep{Asama1996}.
The second and third ways modify the environment.
In the second way, robots construct structures using external materials other than themselves~\citep{Napp2014, Fujisawa2015, Melenbrink2018}.
In the third way, robots construct structures using their own bodies, that is, robots self-assemble structures to overcome the obstacles, such as ramps to pass over a step~\citep{Harada2021}, and bridges to span a gap (e.g., bridges self-assembled from both directions~\citep{Swissler2022}).
In this study, we focus on self-assembled swarm-robot bridges to span a gap, which are generally inspired by living ant bridges self-assembled by ants (more specifically, weaver and army ants)~\citep{Holldobler1978, Franks1989, Holldobler1994}.

We review the related work on self-assembly of bridges to span a gap formed by robots and ants.
In swarm robotics and the neighboring modular robotics, Pamecha et al.~\citep{Pamecha1997} illustrated a concept of self-assembled bridge-like robot structures as a possible application of modular robots, and Hosokawa et al.~\citep{Hosokawa1998} illustrated a concept of self-assembled robot bridges to cross a gap.
Walter et al.~\citep{Walter2002} investigated the self-reconfiguration of hexagonal robots from chains to bridge-like structures by numerical simulations.
Inou et al.~\citep{Inou2000Effect} investigated the influence of mechanical properties on self-assembled bridge-like robot structures to convey a moving load.
Moreover, Inou et al.~\citep{Inou2000Information} examined the information processing functions of robots required for a mission to self-assemble and self-disassemble the structures, and developed an algorithm for the mission.
Furthermore, motion mechanisms for the structures were proposed~\citep{Inou2002c, Inou2003e}.
Later, Suzuki et al.~\citep{Suzuki2011} proposed an algorithm for robots to self-assemble bridge-like structures adaptively in response to load conditions.
Bray and Gro{\ss}~\citep{Bray2021} developed sequential and parallel algorithms for robots to self-assemble cantilevers. Besides, Bray and Gro{\ss}~\citep{Bray2022} developed algorithms for robots to optimally reduce the number of robots belonging to the self-assembled bridges, and to dismantle them.
Nguyen-Duc et al.~\citep{Nguyen-Duc2019} reported the self-assembly and self-disassembly of robot bridges based on a distributed control using numerical simulations.
Malley et al.~\citep{Malley2020} developed a soft-robot system named \textit{Eciton robotica}, and numerically demonstrated that the robots self-assembled and self-disassembled a bridge in response to the local traffic density and the V-shaped gap angle.
Swissler and Rubenstein~\citep{Swissler2022} developed an algorithm for swarm robots to self-assemble amorphous and environment-adaptive three-dimensional structures including cantilevers, and bridges formed from both directions.
Andr{\'{e}}s Arroyo et al.~\citep{AndresArroyo2018} proposed a stochastic algorithm for programmable matter to self-assemble shortcut bridges.
Sugawara et al.~\citep{Sugawara2018} investigated a casualty-based cooperation of swarm robots such that robots overcame a ditch by moving on the dead robots that had unintentionally fallen into the ditch.
Self-assembled floating-robot structures including floating bridges were also proposed~\citep{Paulos2015, Saldana2017b}.
In the research field of the collective behavior of ants, self-assembled living ant bridges were investigated in terms of the robustness and reactivity of bridges~\citep{Garnier2013}, and the dynamical adjustment and optimization of them~\citep{Reid2015}.
Two-dimensional multi-agent systems were developed for reproducing the self-assembly of living ant bridges~\citep{Lutz2018}, and investigating the foraging behavior of ants under the presence of self-assembled living ant bridges to span a gap~\citep{Ishiwata2010, Ichimura2012}.

We also review some related work on the collective robot dynamics in one-dimensional periodic systems. Suematsu et al.~\citep{Suematsu2010} experimentally categorized the collective motion of camphor boats into three types: the homogeneous and congested flows like vehicles, and the cluster flow like ants, and theoretically explained the mechanism of the homogeneous and congested flows. Later, Heisler et al.~\citep{Heisler2012} theoretically explained the mechanism of the cluster flow in detail. Tian et al.~\citep{Tian2018} experimentally found the formation of clusters and the absence of flow-drop like ants, in the collective motion of a kind of self-driven robots (named Hexbug Nano) that did not have any intelligence or pheromone, and theoretically explained the phenomena. These studies focused on the systems without bottleneck. Using a system with a gap will be of interest for better understanding of the collective motion of robots.

\begin{figure}[t]
\centering
\includegraphics[width=\hsize]{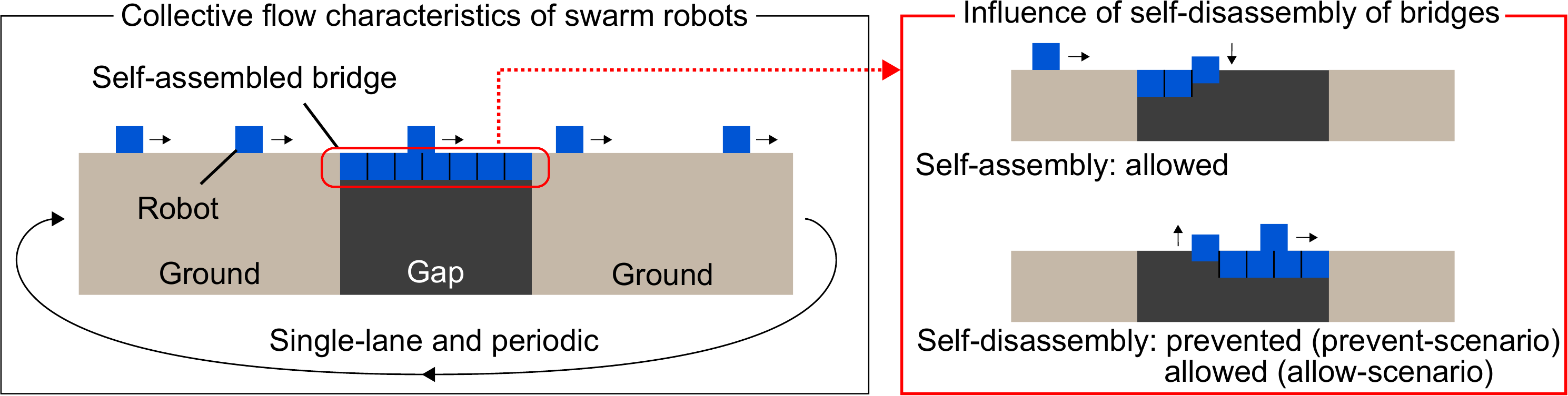}
\caption{Schematic diagram of this study. This study numerically investigates the influence of self-disassembly of bridges on the collective flow characteristics of swarm robots on a single-lane and periodic system with a gap. We use two scenarios: prevent-scenario and allow-scenario. Both the scenarios allow the self-assembly of bridges. Prevent-scenario prevents the self-disassembly of bridges, and allow-scenario allows it.}
\label{fig:schematic_this_study}
\end{figure}
It is well known that the systems in which agents (including robots and ants) self-assemble bridges to span a gap contain trade-offs between the benefit brought by the bridges and the cost caused by them.
For instance, Reid et al.~\citep{Reid2015} demonstrated that army ants dynamically adjusted the positions of the self-assembled living ant bridges to balance the benefit of the increase in the foraging efficiency brought by the bridges against the cost of the decrease in moving ants due to the formation of the bridges.
Andr{\'{e}}s Arroyo et al.~\citep{AndresArroyo2018} produced self-assembled programmable-matter bridges by numerical simulations, and the bridges balanced the benefit of the shortened path length against the cost of the bridges.
In this paper, we consider another cost-benefit trade-off in the swarm robot systems with self-assembled bridges to span a gap: the trade-off in terms of the transport efficiency of swarm robots under the presence of the self-disassembly of bridges.
Self-disassembly of a bridge increases the number of exploring robots, which may increase the transport efficiency of robots.
On the other hand, self-disassembly of a bridge causes the gap to reappear, and disrupts the movement of robots to cross the gap till robots reconstruct the next bridge.
Therefore, self-disassembly of bridges may have significant influence on the collective flow characteristics of swarm robots (such as the flow-density relationship), and we need to clarify the sole effect of the self-disassembly.
As a first step, it is reasonable to simplify the system as much as possible, that is, use a one-dimensional system (such as a single-lane and periodic system with a gap) rather than two or higher dimensional systems~\citep{Ishiwata2010, Ichimura2012, AndresArroyo2018, Lutz2018, Malley2020, Swissler2022}.
Therefore, our aim is to elucidate the influence of self-disassembly of bridges on the collective flow characteristics of swarm robots in a single-lane and periodic system with a gap as shown in Fig.~\ref{fig:schematic_this_study}.
In the system, swarm robots move in one direction on a single-lane road, and pass over the gap by self-assembling a bridge.
To this aim, we consider two scenarios. The first scenario (named prevent-scenario) prevents robots from self-disassembling bridges. In this scenario, robots self-assemble a bridge only once, and the bridge permanently remains. The second scenario (named allow-scenario) allows robots to self-disassemble bridges, and self-assemble new bridges.
We represent the horizontal movement of robots using intelligent driver model (IDM)~\citep{Treiber2000,Treiber2013} as a typical car-following model, and simply model the actions of robots for the self-assembly and self-disassembly of bridges. We investigate the relation between flow and density of swarm robots for the two scenarios and various gap lengths using numerical simulations.
Furthermore, as an additional study, we investigate the dynamics of robots under the absence of the system-periodicity using an open system from the viewpoint of hysteresis.

Since swarm robot is a kind of self-driven particle (SDP), which includes vehicle, pedestrian, insect, and molecular motor, this study contributes to further understanding and development of the collective dynamics of SDPs~\citep{Chowdhury2000, Helbing2001, Schadschneider2010}.

The rest of this paper is organized as follows. We model the system in Sec.~\ref{sec:model}. We show the results in Sec.~\ref{sec:results}, and present the conclusive discussion in Sec.~\ref{sec:discussion}.
\section{\label{sec:model}Model}
We consider a periodic system which has a single-lane road with a gap (such as a deep valley) as shown in Fig.~\ref{fig:initial_state}. The system length is $L$ (m), and the gap length is $\lambda$ (m). We assume that $\lambda$ is a non-negative integer multiple of the length of each robot $d = 0.1\,{\rm m}$, that is, $\lambda = M d$ $(M = 0, 1, 2, \ldots)$. When $\lambda = 0\,{\rm m}$ (that is, $M = 0$), no gap exists in the system.
Position $x$ ({\rm m}) is zero at the right (that is, downstream) edge of the gap, and $L-\lambda$ at the left (that is, upstream) edge of the gap. The range of $x$ is given by $[0, L)$. Let us consider a horizontal movement to the right from $x = 0$ by a distance of $L$. In this movement, $x$ monotonically increases toward $L$ before reaching again the right edge of the gap. When $x$ reaches it, $x$ becomes zero, not $L$.
The system has $N$ robots named robots $1, 2, \ldots, N$. The height of each robot is equal to its length $d$. We define $x_i(t)$ (m) as the position of the front edge of robot $i$ at time $t$ (s), and $v_i(t)$ (${\rm m}/{\rm s}$) as its velocity at time $t$. All robots move to the right  (that is, in the positive $x$ direction). We define density as $\rho = N/L$ (${\rm robots}/{\rm m}$).
In the middle of each run of numerical simulations, no new robot appears in the system, and no robot leaves the system. Hence $N$ and $\rho$ are constant throughout each run.
\subsection{\label{subsec:ini_conds}Initial conditions}
\begin{figure}[t]
\centering
\includegraphics[width=\hsize]{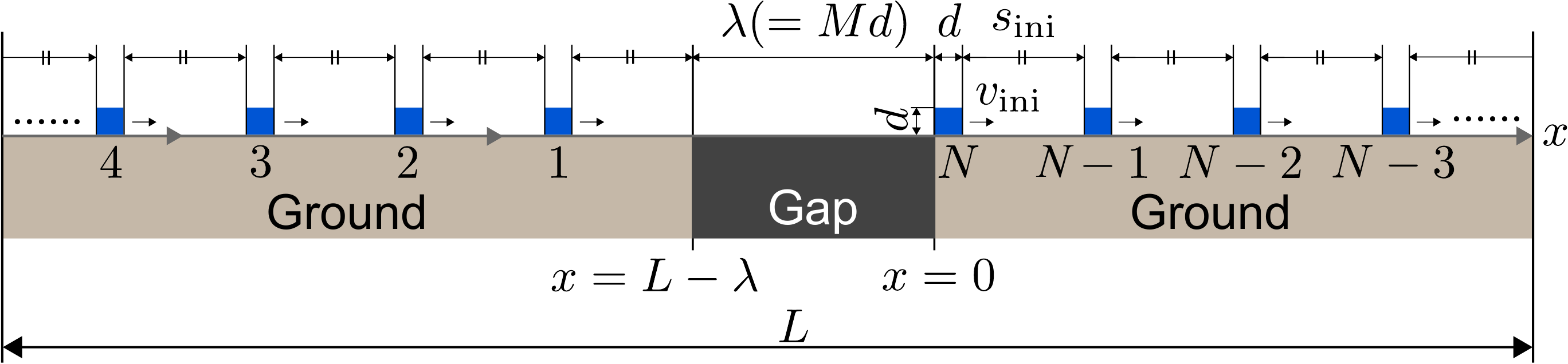}
\caption{Initial conditions of the periodic system with a gap. The system length is $L$. The gap length is $\lambda$ $(= M d)$. The length and height of each robot are $d$. Positions of the left (upstream) and right (downstream) edges of the gap is $x = L - \lambda$ and $x = 0$, respectively. The front edge of robot 1 is initially placed upstream of the left edge of the gap by $s_{\rm ini}$. Robots $2, \ldots, N$ have the same initial forward spacing $s_{\rm ini}$. The rear edge of robot $N$ is initially placed at the right edge of the gap. All robots have the same initial velocity $v_{\rm ini}$, which is the equilibrium velocity of IDM at the forward spacing $s_{\rm ini}$.}
\label{fig:initial_state}
\end{figure}
At the initial time $t = 0\,{\rm s}$, all robots are placed on the continuous section of length $L-\lambda$ as shown in Fig.~\ref{fig:initial_state}.
Robot $N$'s rear edge is initially placed at the right edge of the gap ($x = 0\,{\rm m}$).
Robot $N$ has the initial forward spacing $s_{\rm ini}$ (m).
Robot $N-1$ is initially placed just downstream (ahead) of robot $N$; hence, the rear edge of robot $N-1$ is initially placed at $x = d + s_{\rm ini}$. Besides, robot $N-1$ has the same initial forward spacing $s_{\rm ini}$ as robot $N$.
In the same way, robot $i$'s rear edge is initially placed at $x = (N - i) (d + s_{\rm ini})$ for $i = 1, 2, \ldots, N$, and robots $2, 3, \ldots, N$ have the same initial forward spacing $s_{\rm ini}$.
Additionally, robot $1$'s front edge is initially upstream of the left edge of the gap by $s_{\rm ini}$.
That is, the initial position of the front edge of robot $i$ ($i = 1, \ldots, N$) is given by
\begin{align}
x_{i}(0) = d + (N - i) (d + s_{\rm ini}),
\label{eq:x_i_0}
\end{align}
and $s_{\rm ini}$ is given by
\begin{align}
s_{\rm ini} = \dfrac{L - \lambda}{N} - d.
\label{eq:s_ini}
\end{align}
All robots have the same initial velocity $v_{\rm ini}$ (${\rm m}/{\rm s}$), which is the equilibrium velocity of IDM at the forward spacing $s_{\rm ini}$. We obtain $v_{\rm ini}$ by numerically solving the following equation:
\begin{align}
s_{\rm ini} = \dfrac{s_0 + T v_{\rm ini}}{\sqrt{1 - \left( \dfrac{v_{\rm ini}}{v_0} \right)^{\delta}}}.
\label{eq:s_ini_v_ini}
\end{align}
The definition of IDM is described later in Eq.~(\ref{eq:IDM}), and Eq.~(\ref{eq:s_ini_v_ini}) is obtained by setting $v_i(t) = v_{\rm ini}$, $s_{\rm eff}(t) = s_{\rm ini}$, ${\rm d} v_i(t) / {\rm d} t = 0$, and $\Delta v_{\rm eff}(t) = 0$ in Eq.~(\ref{eq:IDM}).
The parameters of IDM appearing in Eq.~(\ref{eq:s_ini_v_ini}) (that is, $s_0$, $v_0$, $T$, and $\delta$) are explained below Eq.~(\ref{eq:s_asterisk}).
\subsection{\label{subsec:action}Actions of robots}
\begin{figure}[t]
\centering
\includegraphics[width=\hsize]{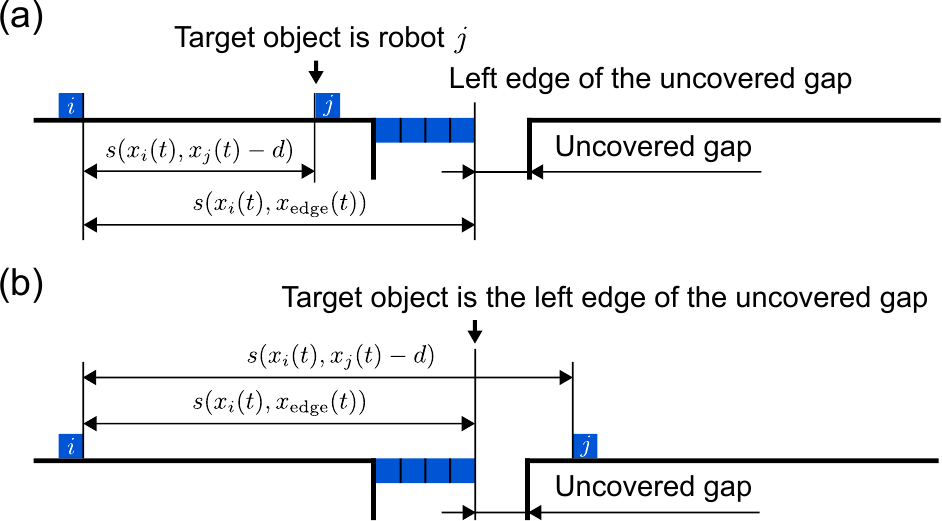}
\caption{
The target object of robot $i$. (a) Forward spacing $s( x_{i}(t), x_{j}(t) - d)$ is smaller than or equal to $s( x_{i}(t), x_{\rm edge}(t))$. The target object of robot $i$ is robot $j$. (b) Forward spacing $s( x_{i}(t), x_{j}(t) - d)$ is greater than $s( x_{i}(t), x_{\rm edge}(t))$. The target object of robot $i$ is the left (upstream) edge of the uncovered gap.}
\label{fig:target_object}
\end{figure}
Each robot has the following functions.
(i) It accelerates and decelerates in the forward direction.
(ii) It detects the forward edge of the road.
(iii) It detects the contact with the ground, walls, and other robots.
(iv) It attaches itself to other robots and walls, and detaches itself from them.
(v) It ascends and descends.
(vi) It communicates with the robots physically connected to itself.
(vii) It obtains its own velocity, the spacing and the relative velocity between it and the object in front of it, and the spacing and the relative velocity between it and the object behind it.
Each robot can perform the three actions using these functions: following the target object, self-assembling a bridge, and self-disassembling it.
\subsubsection{\label{subsubsec:following}Following the target object}
\begin{table}[t]
\caption{
Parameter settings of IDM.
}
\begin{center}
\begin{tabular}{lllllll}
\toprule
$a$ & $b$ & $s_0$ & $v_0$ & $T$ & $\delta$ \\
\midrule
$0.1\,{\rm m}/{\rm s}^2$ & $0.15\,{\rm m}/{\rm s}^2$ & $0.02\,{\rm m}$ & $0.1\,{\rm m}/{\rm s}$ & $1\,{\rm s}$ & 4 \\
\bottomrule
\end{tabular}
\end{center}
\label{table:param}
\end{table}
We define
\begin{subnumcases}{\label{eq:s_x1_x2} s\left( x_1,\,x_2 \right) = }
x_2 - x_1       & \text{if $x_2 \ge x_1$,} \label{eq:s_x1_x2_1} \\
x_2 - x_1 + L & \text{if $x_2 < x_1$,} \label{eq:s_x1_x2_2}
\end{subnumcases}
as the forward spacing from position $x_1$ to position $x_2$.
Robot $i$ moves horizontally to the right by following the target object as shown in Figs.~\ref{fig:target_object}(a) and (b).
A candidate of the target object of robot $i$ is its preceding robot that is not a complete component of a bridge. This preceding robot is named robot $j$.
Another candidate is the left edge of the uncovered gap. The uncovered gap is the part of the gap not covered by the complete bridge components.
As shown in Fig.~\ref{fig:target_object}(a), robot $i$ selects robot $j$ as the target object if
\begin{align}
s \left( x_{i}(t), x_{j}(t) - d \right) \le s \left( x_{i}(t), x_{\rm edge}(t) \right),
\label{eq:cond_target_object}
\end{align}
where position $x_{\rm edge}(t)$ is the position of the left edge of the uncovered gap at time $t$.
As shown in Fig.~\ref{fig:target_object}(b), robot $i$ selects the left edge of the uncovered gap as the target object if
\begin{align}
s \left( x_{i}(t), x_{j}(t) - d \right) > s \left( x_{i}(t), x_{\rm edge}(t) \right).
\label{eq:cond_target_object_2nd}
\end{align}
We define $s_{\rm eff}(t)$ as the forward spacing from robot $i$ to the target object at time $t$:
\begin{subnumcases}{\label{eq:s_eff} s_{\rm eff}(t) = }
s \left( x_{i}(t), x_{j}(t) - d \right)       & \text{if the target object is robot $j$,} \label{eq:s_eff_1} \\
s \left( x_{i}(t), x_{\rm edge}(t) \right) & \text{if the target object is the left edge of the uncovered gap.} \label{eq:s_eff_2}
\end{subnumcases}
Besides, we define $\Delta v_{\rm eff}(t)$ as the relative velocity between robot $i$ and the target object at time $t$:
\begin{subnumcases}{\label{eq:Delta_v_eff} \Delta v_{\rm eff}(t) = }
v_{i}(t) - v_{j}(t)       & \text{if the target object is robot $j$,} \label{eq:Delta_v_eff_1} \\
v_{i}(t) - 0 = v_{i}(t)  & \text{if the target object is the left edge of the uncovered gap.} \label{eq:Delta_v_eff_2}
\end{subnumcases}
The acceleration of robot $i$ at time $t$ is determined by IDM~\citep{Treiber2000,Treiber2013} that uses $s_{\rm eff}(t)$ and $\Delta v_{\rm eff}(t)$:
\begin{align}
\dfrac{{\rm d} v_i(t)}{{\rm d} t} = a \left[ 1 - \left\{ \dfrac{v_i(t)}{v_0} \right\}^{\delta} - \left\{ \dfrac{s^{\ast} \left( v_i(t), \Delta v_{\rm eff}(t) \right) }{s_{\rm eff}(t)} \right\}^2 \right],
\label{eq:IDM}
\end{align}
where
\begin{align}
s^{\ast} \left( v_i(t), \Delta v_{\rm eff}(t) \right) = s_0 + \max \left\{ 0, T v_i(t) + \dfrac{v_i(t) \Delta v_{\rm eff}(t)}{2 \sqrt{ab}} \right\}
\label{eq:s_asterisk}
\end{align}
is the desired 
forward spacing
of robot $i$ at time $t$~\citep{Treiber2013}.
Parameter $a$ (${\rm m}/{\rm s}^2$) is the maximum acceleration, $b$ (${\rm m}/{\rm s}^2$) is the comfortable deceleration, $s_0$ (m) is the forward spacing in the halting state on the ground, $v_0$ (${\rm m}/{\rm s}$) is the desired velocity, $T$ (s) is the safe time spacing, and $\delta$ is the exponent. We set these parameters as listed in Table~\ref{table:param}.

If robot $j$ does not exist and the uncovered gap exists, the target object is the left edge of the uncovered gap.
If robot $j$ exists and the uncovered gap does not exist, the target object is robot $j$.
If neither robot $j$ nor the uncovered gap exists, the target object does not exist, and the acceleration of robot $i$ is given by
\begin{align}
\dfrac{{\rm d} v_i(t)}{{\rm d} t} = a \left[ 1 - \left\{ \dfrac{v_i(t)}{v_0} \right\}^{\delta} \right].
\label{eq:IDM_single_agent}
\end{align}

We update $x_i(t)$ and $v_i(t)$ at regular time intervals of
$\Delta t = 0.01\,{\rm s}$
by
\begin{align}
x_i(t + \Delta t) = x_i(t) + v_i(t) \Delta t + \dfrac{a_i(t) (\Delta t)^2}{2}
\label{eq:x_update_ballistic}
\end{align}
and
\begin{align}
v_i(t + \Delta t) = v_i(t) + a_i(t) \Delta t,
\label{eq:v_update_ballistic}
\end{align}
respectively according to the ballistic method~\citep{Treiber2015}.
Since the system is periodic, if $x_i(t + \Delta t) \ge L$, we subtract $L$ from $x_i(t + \Delta t)$.
\subsubsection{\label{subsubsec:construction}Self-assembling a bridge}
\begin{figure}[t]
\centering
\includegraphics[width=\hsize]{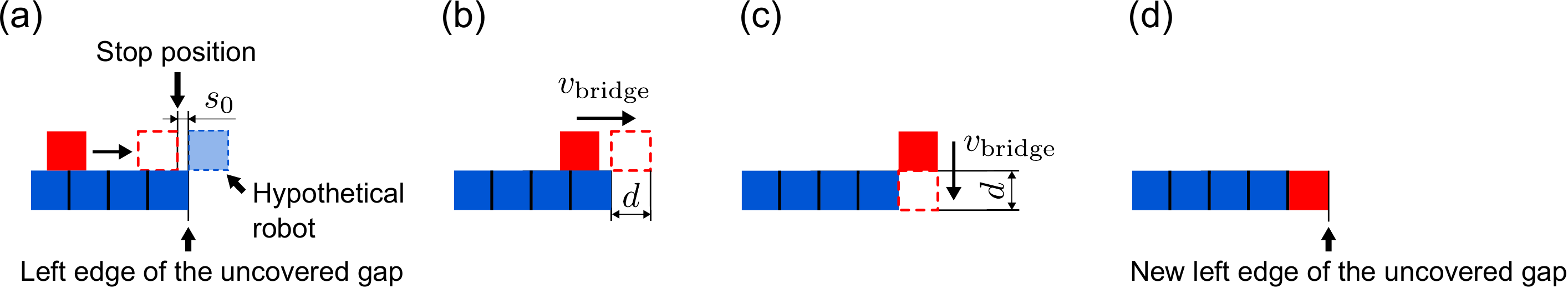}
\caption{
Actions of a robot for self-assembling a bridge. (a) It stops at the position upstream of the left edge of the uncovered gap by $s_0$ as if a hypothetical robot is stopped at the position downstream of the edge by $d$. (b) It moves forward by a distance of $s_0 + d$ at velocity $v_{\rm bridge}$. (c) It descends by a distance of $d$ at velocity $v_{\rm bridge}$.
(d) It becomes a complete component of the bridge at the moment when it has descended by a distance of its height. No time is consumed for bridge connection.
}
\label{fig:actions_self_assembly}
\end{figure}

When the target object of robot $i$ is the left edge of the uncovered gap, robot $i$
stops at the position upstream of the left edge of the uncovered gap by $s_0$ as if a hypothetical robot is stopped at the position downstream of the edge by $d$ as shown in Fig.~\ref{fig:actions_self_assembly}(a).
We judge that robot $i$ has stopped at the stopping position (the position upstream of the edge by $s_0$) if its velocity is smaller than $10^{-3}\,{\rm m}/{\rm s}$, and its position is beyond a point upstream of the stopping position by $10^{-3}\,{\rm m}$.
Next, robot $i$ moves forward by a distance of $s_0 + d$ at velocity $v_{\rm bridge}$ (${\rm m}/{\rm s}$) as shown in Fig.~\ref{fig:actions_self_assembly}(b).
Subsequently, robot $i$ descends by a distance of $d$ at velocity $v_{\rm bridge}$ as shown in Fig.~\ref{fig:actions_self_assembly}(c).
At the moment when robot i has descended by a distance of its height, robot i becomes a complete component of a bridge as shown in Fig.~\ref{fig:actions_self_assembly}(d). No time is consumed for bridge connection.
Self-assembly of a bridge lasts until the bridge length becomes $\lambda$ (that is, $M$ robots become complete components of the bridge).
Robots can move forward on the robots that are complete components of a bridge.

It should be noted that the maximum permissible stress~\citep{Inou2000Effect}, and the maximum permissible moment and axial load~\citep{Bray2021, Bray2022} were considered in self-assembled robot structures.
As a first step to investigate the influence of self-disassembly of bridges on the collective flow characteristics of swarm robots, we do not consider the limitations of permissible stress or moment applied to robots.
Besides, we simplify the bridge structure as much as possible. That is, the bridge is composed of a single layer of robots similarly to the robot-bridge concept illustrated by Hosokawa et al.~\citep{Hosokawa1998}.
Moreover, the single-layer bridges are maintained till robots start self-disassembling them.
\subsubsection{\label{subsubsec:destruction}Self-disassembling a bridge}
\begin{figure}[t]
\centering
\includegraphics[width=0.7\hsize]{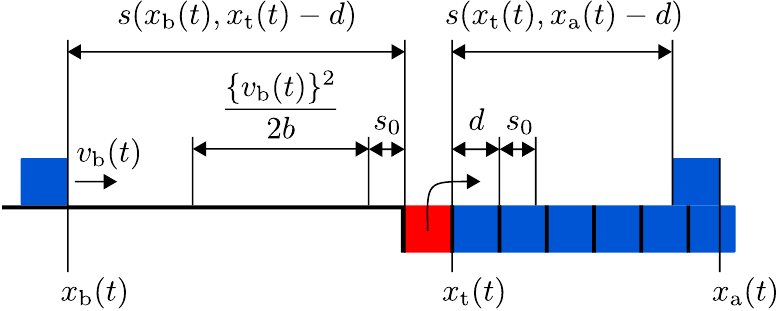}
\caption{Schematic view of the conditions for target-robot to start self-disassembling a bridge.}
\label{fig:conds_self_disassembly}
\end{figure}

\begin{figure}[t]
\centering
\includegraphics[width=0.7\hsize]{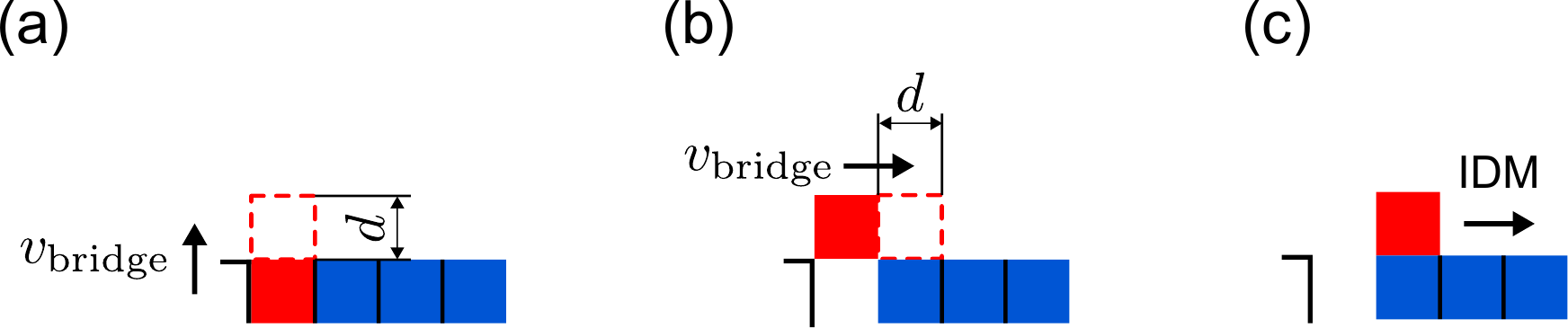}
\caption{Actions of target-robot for self-disassembling a bridge. (a) It ascends by a distance of $d$ at velocity $v_{\rm bridge}$. (b) It moves forward by a distance of $d$ at velocity $v_{\rm bridge}$, and stops. (c) It resumes following its target object according to IDM.}
\label{fig:actions_self_disassembly}
\end{figure}

Self-disassembling a bridge is prevented in prevent-scenario, and allowed in allow-scenario. Therefore, robots perform the following actions only in allow-scenario.
Self-disassembling a bridge is prohibited until self-assembling it is completed (that is, the bridge length becomes $\lambda$).
After the bridge length becomes $\lambda$, the robot that is the most upstream complete component of the bridge is named \textit{target-robot}.
Among the robots that are not complete components of bridges, the robot just downstream of target-robot is named \textit{robot-ahead}, and the robot just upstream of target-robot is named \textit{robot-behind}.
If only one robot is not a complete component of bridges, this robot is robot-ahead and robot-behind simultaneously.
Target-robot checks the following two conditions:
\begin{align}
s(x_{\rm t}(t),  x_{\rm a}(t) - d) \ge d + s_0
\label{eq:conds_destruction_1}
\end{align}
and
\begin{align}
s(x_{\rm b}(t), x_{\rm t}(t) - d) \ge s_0 + \dfrac{ \left\{ v_{\rm b}(t) \right\}^2}{2b},
\label{eq:conds_destruction_2}
\end{align}
as illustrated in Fig.~\ref{fig:conds_self_disassembly}.
Positions $x_{\rm t}(t)$ (m), $x_{\rm a}(t)$ (m), and $x_{\rm b}(t)$ (m) are the positions of target-robot, robot-ahead, and robot-behind at time $t$, respectively. Velocity $v_{\rm b}(t)$ (${\rm m}/{\rm s}$) is the horizontal velocity of robot-behind at time $t$.
Condition~(\ref{eq:conds_destruction_1}) prevents target-robot from colliding with robot-ahead.
In condition (\ref{eq:conds_destruction_2}), $\left\{v_{\rm b}(t)\right\}^2/(2b)$ is the distance required for robot-behind to stop from velocity $v_{\rm b}(t)$ at acceleration $-b$.
Condition~(\ref{eq:conds_destruction_2}) denotes that the forward spacing from robot-behind to target-robot $s(x_{\rm b}(t), x_{\rm t}(t) - d)$ should be greater than or equal to the sum of $s_0$ and this required distance.
If an uncovered gap exists in the forward spacing from robot-behind to target-robot, target-robot does not take robot-behind into account in starting its self-disassembling actions by setting $s(x_{\rm b}(t), x_{\rm t}(t) - d)$ in Cond.~(\ref{eq:conds_destruction_2}) to a sufficiently large value: $s(x_{\rm b}(t), x_{\rm t}(t) - d) = 10^6\,{\rm m}$.
If both the conditions are satisfied, target-robot starts the following self-disassembling actions.
First, target-robot ascends by a distance of $d$ at velocity $v_{\rm bridge}$ as shown in Fig.~\ref{fig:actions_self_disassembly}(a). Second, it moves forward by a distance of $d$ at velocity $v_{\rm bridge}$, and stops as shown in Fig.~\ref{fig:actions_self_disassembly}(b). Third, it resumes moving forward from the velocity of zero by following its target object according to IDM as shown in Fig.~\ref{fig:actions_self_disassembly}(c).
When it starts the third action, it finishes its role as the target robot, and another robot that is the most upstream complete component of the bridge is assigned as the next target-robot.
If neither robot-ahead nor robot-behind exists, target-robot starts the self-disassembling actions without checking Conds.~(\ref{eq:conds_destruction_1}) or (\ref{eq:conds_destruction_2}).
It should be noted that robots are allowed to self-assemble a new bridge while they are self-disassembling a bridge.
\subsection{\label{subsec:measurement_values}Measurement value}
To evaluate the transport efficiency of robots, we measure flow $q$ (${\rm robots}/{\rm s}$) in a time-space rectangular region with the measurement period $T_{\rm meas}$ (s) and the system length $L$ according to Edie's general definition of flow~\citep{Edie1965}.
We define $D$ (${\rm robots} \cdot {\rm m}$) as the total horizontal distance traveled by all the robots in the system within this region.
Flow $q$ is given by $D$ divided by the area of this region $L T_{\rm meas}$~\citep{Edie1965}:
\begin{align}
q = \dfrac{D}{L T_{\rm meas}}.
\label{eq:q}
\end{align}
We set $T_{\rm meas} = 10^4\,{\rm s}$, which is sufficiently large as a measurement period. To obtain $q$ at the steady state, we start measuring $q$ at 
$t = 10^4\,{\rm s}$, and finish at 
$t = 10^4\,{\rm s} + T_{\rm meas}$. We name $q$ in prevent-scenario and allow-scenario $q_{\rm prev}$ and $q_{\rm allow}$, respectively.
\section{\label{sec:results}Results}
\begin{figure}[t]
\centering
\includegraphics[width=\hsize]{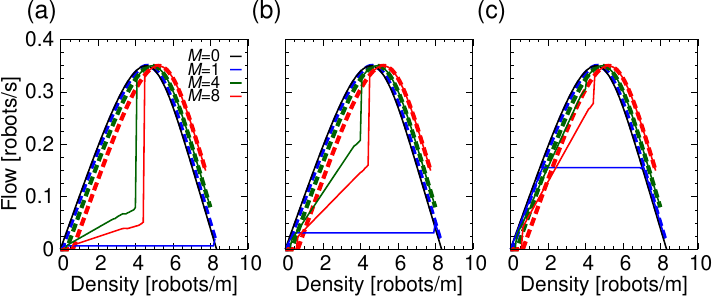}
\caption{Flow--density diagrams for 
$L=12\,{\rm m}$. (a) $v_{\rm bridge} = 0.002\,{\rm m}/{\rm s}$. (b) $v_{\rm bridge} = 0.01\,{\rm m}/{\rm s}$. (c) $v_{\rm bridge} = 0.05\,{\rm m}/{\rm s}$.
The thick dashed blue, dark-green, and red lines denote those for prevent-scenario at $M=1$, 4, and 8, respectively.
The thin solid blue, dark-green, and red lines denote those for allow-scenario at $M=1$, 4, and 8, respectively.
The thin solid black lines denote those for the scenario with no gap ($M=0$).}
\label{fig:flow-density_diagrams}
\end{figure}
\begin{figure}[htbp]
\centering
\includegraphics[width=\hsize]{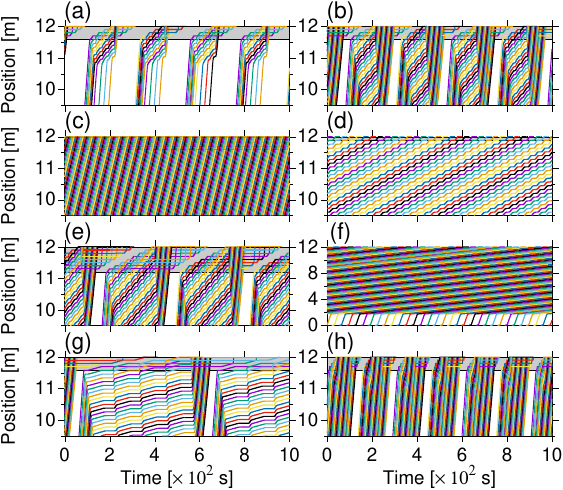}
\caption{Time--space diagrams for allow-scenario. We set 
$L=12\,{\rm m}$. We shift the display time from the true time by 
$-10^4\,{\rm s}$; therefore, the time zero displayed in these diagrams denotes 
$t = 10^4\,{\rm s}$. Gray regions denote gaps as guides to the eyes.
(a)--(c) $N = 12$, 36, and 84, respectively. $M = 4$. $v_{\rm bridge} = 0.01\,{\rm m}/{\rm s}$.  (d) and (e) $M = 1$ and 8, respectively. $N = 36$. $v_{\rm bridge} = 0.01\,{\rm m}/{\rm s}$. (f) $M = 1$. $N = 84$. $v_{\rm bridge} = 0.01\,{\rm m}/{\rm s}$. (g) and (h) $v_{\rm bridge} = 0.002\,{\rm m}/{\rm s}$ and $0.05\,{\rm m}/{\rm s}$, respectively. $N = 36$. $M = 4$.
}
\label{fig:time-space_diagrams}
\end{figure}
\begin{figure}[t]
\centering
\includegraphics[width=0.67\hsize]{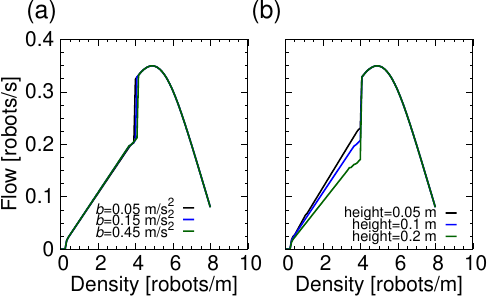}
\caption{
(a) Flow--density diagrams for $b = 0.05$, $0.15$, and $0.45\,{\rm m}/{\rm s^2}$.
(b) Flow--density diagrams for robot height of 0.05, 0.1, and $0.2\,{\rm m}$.
We set allow-scenario with $M = 4$ and $v_{\rm bridge} = 0.01\,{\rm m}/{\rm s}$. The other parameters are same as in Fig.~\ref{fig:flow-density_diagrams}(b).
}
\label{fig:flow-density_b_height}
\end{figure}
\begin{figure}[t]
\centering
\includegraphics[width=\hsize]{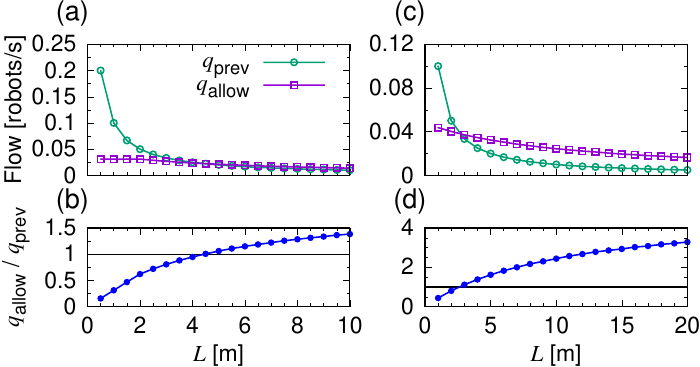}
\caption{
(a) and (c) Flow in prevent-scenario $q_{\rm prev}$, and flow in allow-scenario $q_{\rm allow}$ as functions of the system length $L$.
(b) and (d) Ratio $q_{\rm allow} / q_{\rm prev}$ as a function of $L$. The solid black lines denote ratio of 1 as guides to the eyes.
We set $N = 2$ and $M = 1$ for (a) and (b), and $N = 5$ and $M = 4$ for (c) and (d).
We set $v_{\rm bridge} = 0.01\,{\rm m}/{\rm s}$.
The solid lines through data points are guides to the eyes.
}
\label{fig:q_allow_q_prev}
\end{figure}
To reveal the relation between flow and density, we have conducted numerical simulations for 
$L = 12\,{\rm m}$, $M = \lambda/d \in \left\{0, 1, 4, 8\right\}$, $v_{\rm bridge} \in \left\{ 0.002\,{\rm m}/{\rm s}, 0.01\,{\rm m}/{\rm s}, 0.05\,{\rm m}/{\rm s} \right\}$, and $N \in \left\{ 1, 2, \ldots, N_{\rm max}(\lambda) \right\}$.
Parameter $N_{\rm max}(\lambda)$ is the maximum number of robots as a function of $\lambda$, and is given by
\begin{align}
N_{\rm max}(\lambda) = \left\lfloor  \dfrac{L - \lambda}{s_0 + d} \right\rfloor,
\label{eq:N_max_lambda}
\end{align}
where $\left\lfloor \cdot \right\rfloor$ is the floor function such that $\left\lfloor x \right\rfloor = \max \left\{ n \in \mathbb{Z} \mid n \le x \right\}$ for $x \in \mathbb{R}$.
Figures~\ref{fig:flow-density_diagrams}(a)--(c) show flow--density ($q$--$\rho$) diagrams for (a) $v_{\rm bridge} = 0.002\,{\rm m}/{\rm s}$, (b) $v_{\rm bridge} = 0.01\,{\rm m}/{\rm s}$, and (c) $v_{\rm bridge} = 0.05\,{\rm m}/{\rm s}$.
We depict the flow-density diagrams for prevent-scenario at $M=1$, 4, and 8 with the thick dashed blue, dark-green, and red lines, respectively.
We depict those for allow-scenario at $M=1$, 4, and 8 with the thin solid blue, dark-green, and red lines, respectively.
We depict those for the scenario with no gap ($M=0$) with the thin solid black lines.
Figures~\ref{fig:time-space_diagrams}(a)--(h) show time--space diagrams for allow-scenario.
Unless otherwise specified, we set 
$N = 36$, $M = 4$, and $v_{\rm bridge} = 0.01\,{\rm m}/{\rm s}$ for these time-space diagrams.
Figures~\ref{fig:time-space_diagrams}(a)--(c)
show those for 
$N = 12$, 36, and 84, respectively.
Figures~\ref{fig:time-space_diagrams}(d) and (e)
show those for $M = 1$ and 8, respectively.
Figure~\ref{fig:time-space_diagrams}(f) shows that for $N = 84$
and $M = 1$.
Figures~\ref{fig:time-space_diagrams}(g) and (h) show those for $v_{\rm bridge} = 0.002\,{\rm m}/{\rm s}$ and $0.05\,{\rm m}/{\rm s}$, respectively.

First, we focus on the scenario with no gap. As shown in Figs.~\ref{fig:flow-density_diagrams}(a)--(c), there is no sharp drop in flow $q$. The scenario with no gap exhibits only the free-flow steady state, and does not exhibit the jammed steady state, which is frequently observed in the collective dynamics of SDPs~\citep{Chowdhury2000, Helbing2001, Schadschneider2010}. This is because the traffic flow of robots is stable under the parameter settings of IDM.

Second, we focus on prevent-scenario. As shown in Figs.~\ref{fig:flow-density_diagrams}(a)--(c), prevent-scenario also exhibits only the free-flow steady state.
As $M$ increases, flow--density diagrams shift to the higher-density region. This shift occurs because the increase in $M$ increases the number of robots becoming bridge components, and decreases the number of robots moving horizontally.

Third, we focus on allow-scenario. In contrast to the scenario with no gap and prevent-scenario, allow-scenario exhibits two steady states in the ascending order of density: the state of repeated self-assembly and self-disassembly of bridges (named the \textit{assembling-disassembling} state) as shown in Figs.~\ref{fig:time-space_diagrams}(a) and (b), and the free-flow state as shown in Fig.~\ref{fig:time-space_diagrams}(c).
We define $\rho_{\rm cr,\,AF}$ as the critical density separating the assembling-disassembling and free-flow steady states.

We compare flow--density diagrams for allow-scenario with those for prevent-scenario (Figs.~\ref{fig:flow-density_diagrams}(a)--(c)) in the following density regions: (i) $0 < \rho < \rho_1$, (ii) $\rho_1 < \rho < \rho_{\rm cr,\,AF}$, and (iii) $\rho_{\rm cr,\,AF} < \rho \le N_{\rm max}(\lambda)/L$.
Density $\rho_1$ is the threshold density for the order of the magnitude relation between $q_{\rm prev}$ and $q_{\rm allow}$, and is much lower than $\rho_{\rm cr,\,AF}$.
(i) When $\rho$ is extremely low ($0 < \rho < \rho_1$), $q_{\rm allow}$ is greater than $q_{\rm prev}$ for some parameter settings such as $M \in \left\{4, 8\right\}$ and $v_{\rm bridge} \in \left\{0.01\,{\rm m}/{\rm s}, 0.05\,{\rm m}/{\rm s} \right\}$.
We investigate $q_{\rm allow}$ becoming greater than $q_{\rm prev}$ in detail in the later part of this section.
(ii) For $\rho_1 < \rho < \rho_{\rm cr,\,AF}$, $q_{\rm allow}$ is smaller than $q_{\rm prev}$ because repeated self-assembly and self-disassembly of bridges disrupt the horizontal movement of robots as shown in Figs.~\ref{fig:time-space_diagrams}(a), (b), (d), (e), (f), (g), and (h).
(iii) For $\rho_{\rm cr,\,AF} < \rho \le N_{\rm max}(\lambda)/L$, self-disassembly of bridges no longer occurs in the steady state of allow-scenario as shown in Figs.~\ref{fig:time-space_diagrams}(c), and $q_{\rm allow}$ agrees with $q_{\rm prev}$.

We investigate the influence of $M$ on $q_{\rm allow}$ in the assembling-disassembling state (Figs.~\ref{fig:flow-density_diagrams}(a)--(c)).
Flow $q_{\rm allow}$ in this state for $M = 1$ is constant and much smaller than those for $M = 4$ and 8 for a wide density region. Besides, $\rho_{\rm cr,\,AF}$ for $M = 1$ is much greater than that for $M = 4$ and 8. In the allow-scenario with $M = 1$, robots becoming a bridge tends to start disassembling the bridge before the following robot moves on the bridge, as shown in
Fig.~\ref{fig:time-space_diagrams}(d).
This disassembling action lasts permanently, causes a fixed congested area upstream of the gap as shown in Fig.~\ref{fig:time-space_diagrams}(f), and strongly restricts the traffic of robots crossing the gap in a wide range of density.  
Flow $q_{\rm allow}$ in this state for $M = 4$ tends to be larger than that for $M = 8$. This tendency occurs because robots wait for a longer period from joining a bridge to leaving it as $M$ increases, as shown in the time-space diagrams under $v_{\rm bridge} = 0.01\,{\rm m}/{\rm s}$ for $M = 4$ and 8  (Figs.~\ref{fig:time-space_diagrams}(b) and (e), respectively).  Especially, when $\rho$ is sufficiently small (such as $\rho$ roughly lower than $2\,{\rm robots}/{\rm m}$ for $v_{\rm bridge} = 0.05\,{\rm m}/{\rm s}$), $q_{\rm allow}$ in this state decreases monotonically as $M$ increases from $M = 1$ to 4 and 8. 

We investigate the influence of $v_{\rm bridge}$ on $q_{\rm allow}$ in the assembling-disassembling state (Figs.~\ref{fig:flow-density_diagrams}(a)--(c)). Flow $q_{\rm allow}$
in this state increases monotonically
as $v_{\rm bridge}$ increases from $v_{\rm bridge} = 0.002$ to 0.01 to $0.05\,{\rm m}/{\rm s}$. The bottleneck effect caused by the repetition of self-assembly and self-disassembly of bridges becomes weaker as $v_{\rm bridge}$ increases, as shown in the time-space diagrams under $M = 4$ for $v_{\rm bridge} = 0.002$, 0.01, and $0.05\,{\rm m}/{\rm s}$ (Figs.~\ref{fig:time-space_diagrams}(g), (b), and (h), respectively).

We also investigate the influence of the parameters related to assembling and/or disassembling bridges other than $v_{\rm bridge}$ on flow-density diagrams. Such related parameters are IDM parameter $b$, and robot height.
Unless otherwise specified, we use allow-scenario with $M = 4$ and $v_{\rm bridge} = 0.01\,{\rm m}/{\rm s}$, and set the other parameters to the same values as in Figs.~\ref{fig:flow-density_diagrams}(b).
Figure~\ref{fig:flow-density_b_height}(a) shows the flow-density diagrams for three values of $b$ ($b = 0.05$, 0.15, and $0.45\,{\rm m}/{\rm s^2}$). Since we assume that robots can decelerate quickly, we do not set $b$ to too small values. Parameter $b$ influences the start of the self-disassembling action. As $b$ increases, the robot being the uppermost bridge component is more likely to start disassembling, and the state is more likely to become the assembling-disassembling state. Figure~\ref{fig:flow-density_b_height}(a) shows that increasing $b$ slightly enlarges the density region of the assembling-disassembling state. However, since $b$ in this range is considerably high, $b$ in this range does not have significant effect on the flow-density relationship.
Figure~\ref{fig:flow-density_b_height}(b) shows the flow-density diagrams for three values of robot height (0.05, 0.1, and $0.2\,{\rm m}$) under fixed robot length of $0.1\,{\rm m}$. Robot height influences the descending and ascending periods required for robots to complete in the self-assembling and self-disassembling actions, respectively. As robot height increases, it takes more time for robots to complete in descending and ascending actions, which leads to more strong bottleneck effect of bridge construction and destruction. Figure~\ref{fig:flow-density_b_height}(b) shows that increasing robot height indeed decreases flow in the assembling-disassembling state.

Next, we compare $q_{\rm allow}$ with $q_{\rm prev}$ for various system lengths $L$ under fixed $N$ and $M$. We set $v_{\rm bridge} = 0.01\,{\rm m}/{\rm s}$.
Figures~\ref{fig:q_allow_q_prev}(a) and (c) show $q_{\rm prev}$ and $q_{\rm allow}$ as functions of $L$ for (a) $N = 2$ and $M = 1$, and (c) $N = 5$ and $M = 4$.
Figures~\ref{fig:q_allow_q_prev}(b) and (d) show ratio $q_{\rm allow}/q_{\rm prev}$ as a function of $L$ for the same scenarios as in Figs.~\ref{fig:q_allow_q_prev}(a) and (c), respectively.

First, we focus on the case of $N = 2$ and $M = 1$ (Figs.~\ref{fig:q_allow_q_prev}(a) and (b)).
Both $q_{\rm prev}$ and $q_{\rm allow}$ decrease with respect to $L$.
Ratio $q_{\rm allow}/q_{\rm prev}$ is smaller than 1 from $L = 0.5\,{\rm m}$ to nearly $4.5\,{\rm m}$ because the repetition of self-assembly and self-disassembly of bridges disrupts the horizontal movement of robots.
Ratio $q_{\rm allow}/q_{\rm prev}$ increases with respect to $L$.
This increase in $q_{\rm allow}/q_{\rm prev}$ occurs for the following two reasons.
First, as $L$ becomes larger under fixed $N$ and $M$, the arrival rate of robots (${\rm robots} / {\rm s}$) at the gap decreases; therefore, the repetition of self-assembly and self-disassembly of bridges is less likely to disrupt the horizontal movement of robots. Second, self-disassembly of bridges increases the number of robots moving horizontally.
Ratio $q_{\rm allow}/q_{\rm prev}$ is greater than 1 for nearly $L > 4.5\,{\rm m}$, and reaches 1.38 at $L = 10\,{\rm m}$.
Thus, self-disassembly of bridges can increase the transport efficiency of the system when $L$ is sufficiently large for fixed $N$ and $M$.

Second, we focus on the case of $N = 5$ and $M = 4$ (Figs.~\ref{fig:q_allow_q_prev}(c) and (d)).
Both $q_{\rm prev}$ and $q_{\rm allow}$ decrease with respect to $L$.
When $L = 1\,{\rm m}$, which is the minimum value of $L$ to initially place five robots on the ground (the gap length $Md = 0.4\,{\rm m}$, and $N(d + s_0) = 0.6\,{\rm m}$), ratio $q_{\rm allow}/q_{\rm prev}$ is 0.44.
Ratio $q_{\rm allow}/q_{\rm prev}$ increases with respect to $L$ for $L \ge 1\,{\rm m}$, is greater than 1 for roughly $L > 3\,{\rm m}$, and reaches 2.44 for $L = 10\,{\rm m}$, and 3.29 for $L = 20\,{\rm m}$.
Hence ratio $q_{\rm allow}/q_{\rm prev}$ for $N = 5$ and $M = 4$ can be greater than that for $N = 2$ and $M = 1$.

\begin{figure}[t]
\centering
\includegraphics[width=0.5\hsize]{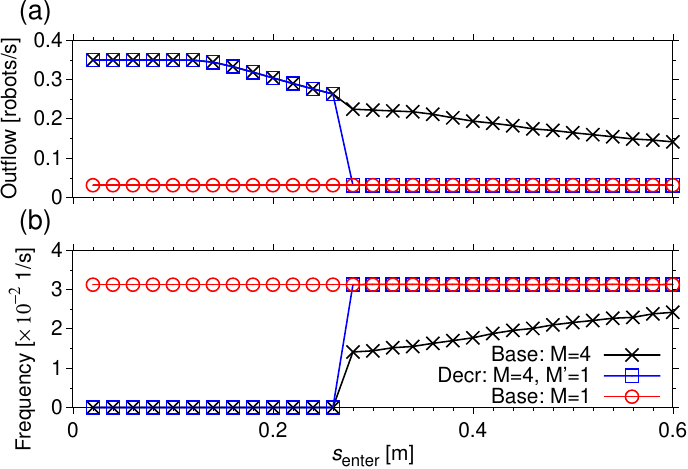}
\caption{
(a) Outflows as
functions
of spatial gap $s_{\rm enter}$ in an open system.
(b) Frequencies of starting self-disassembling actions as
functions
of $s_{\rm enter}$.
Abbreviations ``Base: $M=4$'' and ``Base: $M=1$'' denote baseline-scenario with $M=4$ and $M=1$, respectively. Abbreviation ``Decr: $M=4$, $M'=1$'' denotes decreasing-$M$-scenario with $M=4$ and $M'=1$.
We set allow-scenario with $v_{\rm bridge} = 0.01\,{\rm m}/{\rm s}$.
The lines
through points are guides to the eyes.
}
\label{fig:flow_spatial_gap}
\end{figure}
It should be noted that the density dependence of the presence/absence of the assembling-disassembling state as shown in Figs.~\ref{fig:flow-density_diagrams} and \ref{fig:time-space_diagrams} is strongly influenced by the periodic boundary conditions and the steady-state conditions. Hereafter, we check the dynamics of robots under the absence of the periodic boundary conditions. We focus on seeking the presence of hysteresis in an open system.

We construct our open system as follows. The system has inflows of a regularly spatial gap of $s_{\rm enter}$ (m) . The system length is $10\,{\rm m}$. The system has the left (that is, upstream) boundary at $x = 0\,{\rm m}$, the right (that is, downstream) boundary at $x = 10\,{\rm m}$, and a gap with a length of $M d$ at $5 - M d \le x  \le 5\,{\rm m}$. At the initial time $t = 0\,{\rm s}$, the first robot appears at $x = 0\,{\rm m}$ with its initial velocity, which is the equilibrium velocity of IDM at spatial gap $s_{\rm enter}$. Robots move to the right (that is, downstream) inside the system, and cross the gap by self-assembling and/or disassembling bridges with $v_{\rm bridge} = 0.01\,{\rm m}/{\rm s}$. The following robots enter the system near the left boundary with spatial gap
$s_{\rm enter}$ and their initial velocity, which is the equilibrium velocity at spatial gap $s_{\rm enter}$.
Robots are removed from the system when they go beyond the right boundary.

To check the presence of hysteresis, we use two scenarios for self-disassembly rules. 
In the first scenario (we call it baseline-scenario),  robots self-disassemble bridges according to the rules described in Sec.~\ref{subsubsec:destruction}. 
In the second scenario (we call it decreasing-$M$-scenario), robots can self-disassemble bridges similarly to baseline-scenario from the initial time $t = 0\,{\rm s}$ to $t = 10^4\,{\rm s}$. After $t = 10^4\,{\rm s}$, we permit only $M'$ bridge components (from the most upstream component to the $M'$-th upstream component) to perform self-disassembling actions. The other $M - M'$ components (from $(M' + 1)$-th upstream component to the most downstream component) are prevented from performing self-disassembling actions.
We set two output values. The first value is the outflow of robots (${\rm robots}/{\rm s}$) from the system. The second one is the frequency of starting self-disassembling actions ($1/{\rm s}$) in the system. We measure these values for a period of $10^4\,{\rm s}$ after a preparation time of $10^4\,{\rm s}$. More precisely, we measure them for $10^4 \le t \le 2 \times 10^4\,{\rm s}$ in baseline-scenario, and for $2 \times 10^4 \le t \le 3 \times 10^4\,{\rm s}$ in decreasing-$M$-scenario. 
 
We perform numerical simulations for the two scenarios in the open system. We set $M \in \left\{1, 4\right\}$ for baseline-scenario, and $M = 4$ and $M' = 1$ for decreasing-$M$-scenario. We use only allow-scenario regarding self-disassembly. The other unmentioned parameters are same as in Fig~\ref{fig:flow-density_diagrams}(b).
Figures~\ref{fig:flow_spatial_gap}(a) and (b)
shows outflows as
functions
of $s_{\rm enter}$, and frequencies of starting self-disassembling actions as
functions
of $s_{\rm enter}$, respectively.
As shown in Fig.~\ref{fig:flow_spatial_gap}(a), when $s_{\rm enter} \ge 0.28\,{\rm m}$, outflow in decreasing-$M$-scenario agrees with that in baseline-scenario with $M = 1$. When $s_{\rm enter} \le 0.26\,{\rm m}$, outflow in decreasing-$M$-scenario is much higher than that in baseline-scenario with $M = 1$, and agrees with that in baseline-scenario with $M = 4$.
As shown in Fig.~\ref{fig:flow_spatial_gap}(b), when $s_{\rm enter} \ge 0.28\,{\rm m}$, frequency of starting self-disassembling actions in decreasing-$M$-scenario agrees with that in baseline-scenario with $M = 1$. When $s_{\rm enter} \le 0.26\,{\rm m}$, the frequency in decreasing-$M$-scenario is zero, and agrees with that in baseline-scenario with $M = 4$.
These results show that there is a hysteresis in the open system. Setting first $M = 4$, and subsequently shifting to $M'=1$ maintains the state of $M = 4$ only when $s_{\rm enter}$ is sufficiently small. The dynamics of robots does not require the periodic boundary conditions to exhibit the hysteresis.   
\section{\label{sec:discussion}Discussion}
We have revealed the following phenomena by numerical simulations.
Flow-density diagrams shift to the higher-density region as the gap length becomes larger for both the scenarios preventing self-disassembly of bridges (prevent-scenario) and allowing it (allow-scenario).
When density of robots $\rho$ is low, the steady state of repeated self-assembly and self-disassembly of bridges (the assembling-disassembling state) emerges in allow-scenario. Flow in allow-scenario $q_{\rm allow}$ in the assembling-disassembling state is greater than flow in prevent-scenario $q_{\rm prev}$ if $\rho$ is extremely low, or the system length $L$ is sufficiently large under fixed values of the number of robots $N$ and the gap-length parameter (the gap length divided by the robot length) $M$. Otherwise, $q_{\rm allow}$ in this state is smaller than $q_{\rm prev}$. Flow $q_{\rm allow}$ in this state increases monotonically with respect to the velocity of robots during joining and leaving bridges $v_{\rm bridge}$.
Our results suggest that self-disassembly of bridges in periodic systems is recommended in terms of the transport efficiency of robots only if $\rho$ is extremely low.
Furthermore, as an additional research, we have found that the dynamics of robots exhibits hysteresis under the absence of periodic boundaries, that is, in an open system.
Our findings contribute to the development of the collective dynamics of SDPs that self-assemble and self-disassemble structures, and pave the way for elucidating the collective dynamics with other types of self-assembled structures (e.g., ramps~\citep{Harada2021}, chains~\citep{Swissler2022}, and towers~\citep{Cucu2015, Swissler2022}).

Flow restriction under low-density conditions occurs in not only our system in the assembling-disassembling state, but also the unidirectional, single-lane, and periodic ant traffic system represented by a cellular automaton model~\citep{Chowdhury2002, Schadschneider2010}.
In the ant traffic system, ants attach chemical substance called pheromone to the ground. The probability that an ant moves forward is high or low if the pheromone exists or does not exist just ahead of the cell where the ant is placed, respectively. Flow restriction is caused in the ant traffic system by the evaporation of the pheromone.
Both the ant traffic system and our system in allow-scenario have common in that flow restriction is mitigated by a sufficiently large density, whereas the two systems are different with respect to the site-specificity of the bottleneck effect.
In the ant traffic system, evaporation of the pheromone occurs irrespective of sites. In our system, self-assembly and self-disassembly of bridges are site-specific, and occur at the gap.

In this study, a single type of robots have played multiple roles: moving horizontally to provide transportation, and being static to provide bridge infrastructure. Using only a single type of robots is not a unique way to realize the transportation and infrastructure. Using a type of robots for transportation and another type of
robots
for bridges would be beneficial for the system efficiency. For instance, in the work of Inou et al.~\citep{Inou2000Effect}, robots self-assembled bridge-like structures to produce a road for a moving load that was illustrated differently from the bridge robot. Paulos et al.~\citep{Paulos2015} and Salda{\~{n}}a et al.~\citep{Saldana2017b} investigated the floating robots self-assembling floating bridges and not moving on the bridges. Investigating the transportation efficiency using multiple types of specialized robots will be investigated in our future work.

We list some potential future work as follows.
Our model has treated single-layer bridges, and has not considered the specifications of robots in detail:
(i) the maximum permissible stress, moment or axial force~\citep{Inou2000Effect, Bray2021, Bray2022},
(ii) physical interfaces to attach themselves to other robots~\citep{Swissler2020} or walls,
(iii) motion mechanisms for self-assembling the structures~\citep{Inou2002c, Inou2003e},
(iv) horizontal movement mechanisms, such as tracks or wheels~\citep{Mondada2004, Cucu2015, Harada2021},
(v) sensors to detect joint positions~\citep{Ostergaard2006b} or terrain changes~\citep{OGrady2010a},
or (vi) wireless communication devices~\citep{Tang2009c}.
We will take these specifications into account for the systems with self-assembled bridges of more complex structures~\citep{Inou2000Effect, Bray2021, Bray2022} in our future work.
We have focused on a
unidirectional and single-lane
system. We will treat other types of systems, such as
bidirectional and/or multi-lane systems.
Since self-assembly and self-disassembly of bridges cause jamming clusters of robots, removing the clusters is expected to increase the transport efficiency of robots.
Removal of the clusters will be achieved by the strategies developed in vehicular traffic flow such as the jam-absorption driving~\citep{Nishi2013, Taniguchi2015}. To introduce the strategies into the swarm robot system warrants our future work.




\end{document}